\documentclass[pra, showpacs, twocolumn, floatfix]{revtex4}
\usepackage{graphicx}
\usepackage{amsmath, amsfonts, amssymb, bm}
\begin{document}
\title{Collectively enhanced resonant photoionization in a multi-atom ensemble}
\author{Carsten M\"uller}
\email{c.mueller@mpi-hd.mpg.de}
\author{Mihai A. \surname{Macovei}}
\email{mihai.macovei@mpi-hd.mpg.de}
\author{Alexander B. Voitkiv}
\email{a.voitkiv@mpi-hd.mpg.de}
\affiliation{Max-Planck-Institut f\"{u}r Kernphysik, Saupfercheckweg
1, D-69117 Heidelberg, Germany}
\date{\today}
\begin{abstract}
Photoionization of an atom via interatomic correlations to $N$ neighboring atoms may be strongly enhanced due to constructive interference of quantum pathways. The ionization proceeds via resonant photoexcitation of a neighbor atom and subsequent interatomic Coulombic decay. The enhancement can scale with $N^2$, leading to ``super-enhanced photoionization''.
\end{abstract}
\maketitle

The quantum dynamics of a single fluorescing atom is strongly modified 
when an additional emitter is located nearby \cite{Dck}. Furthermore, 
it is well known that collective interactions between closely spaced 
particles can lead to a significant modification of spontaneous 
emission processes - an effect called superradiance. In particular, the 
quantum dynamics of an excited collective system can be $N$ times faster than for a 
single particle, and the intensity of the emitted electromagnetic field scales 
as $N^{2}$ in multiparticle samples, where $N$ is the number of radiators. 
The enhancement occurs due to constructive interference effects. Superradiance 
was first demonstrated experimentally in an optically pumped 
hydrogen fluoride gas \cite{exp} and the concept was transferred to other 
branches of physics \cite{bran}, including sonoluminescence \cite{sono},
plasmas \cite{plas}, Bose-Einstein condensates \cite{bec}, molecular nanomagnets 
\cite{moln}, gamma ray lasers \cite{bald}, and band gap materials \cite{bgm}.

The process of photoionization may be influenced by the atomic environment as well. Indeed, when photoionization of an atom $A$ occurs in the presence of a neighboring atom $B$, then -- apart from the direct ionization of $A$ without participation of $B$ -- there are additional resonant channels for certain photon energies where first $B$ is photoexcited and afterwards, upon deexcitation, $B$ transfers the transition energy radiationlessly to $A$ leading to its ionization. The latter Auger-like step is commonly known today as interatomic Coulombic decay (ICD) \cite{ICD,ICDexp,ICDres,ICDrev}. The effect of the environment on photoionization has been observed on core resonances in metall oxides, where it was termed Multi-Atom Resonant Photo\-emission (MARPE) \cite{MARPE}. The measured MARPE effects, amounting to about 30--100\,\% on the resonance and 10--30\,\% after energy integration, were substantial but did not exceed the direct photoionization.

Recently, it has been shown that -- in certain circumstances -- the interatomic channel of photoionization via ICD can be very strong and by far the dominant one \cite{2CPI}. In fact, its strength with respect to the direct channel is given by the ratio \cite{2CPI,2CDR}
\begin{eqnarray}
\rho_1 = \frac{\sigma_{\rm interatomic}^{(N=1)}}{\sigma_{\rm direct}}\sim \left(\frac{c}{\omega R} \right)^6\,,
\label{rho1}
\end{eqnarray}
where $\omega$ denotes the electron transition frequency, $R$ the interatomic distance and $c$ the speed of light. Equation~(\ref{rho1}) assumes that the bound-bound transition is dipole-allowed and the radiative width in atom $B$ exceeds the interatomic Auger width. The latter will always hold at sufficiently large values of $R$. When the resonance frequency is not too large, $\omega\lesssim 1$--10\,eV, the ratio $\rho_1$ exceeds unity for interatomic distances $R\lesssim 1$--10\,nm.

In the present note we show that the strength of the interatomic channel of photoionization can be largely enhanced even further when atom $A$ is surrounded by $N>1$ atoms $B$ (see Fig.\,1). In this case, the photoionization amplitude must be coherently summed over all neighbor atoms, allowing for collective effects via interfering quantum pathways. Under certain conditions, the resulting enhancement can be proportional to $N^2$ due to fully constructive interference. Conditions under which this effect of super-enhanced photoionization may occur, will be specfied. We also outline the prospects for an experimental observation of collective enhancements in multi-atom resonant photoionization.

\begin{figure}[b]  
\vspace{-0.25cm}
\begin{center}
\includegraphics[width=0.4\textwidth]{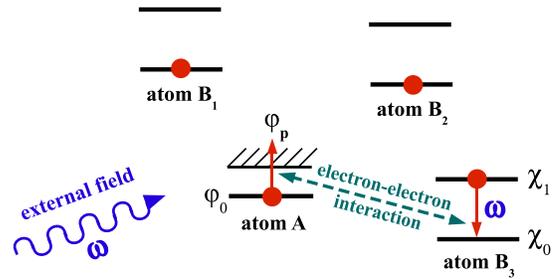}
\end{center}
\vspace{-0.5cm} 
\caption{Scheme of photoionization of atom $A$ in the presence of an external laser field and $N=3$ neighboring atoms $B_j$ ($j=1,2,3$). Apart from the direct photoionization of $A$ there are interatomic channels via photoexcitation of one of the atoms $B_j$ (here $j=3$) and subsequent ICD. }
\label{figure1}
\end{figure}

Our theoretical description of photoionization of atom $A$ in the presence of an electromagnetic wave and $N$ neighboring atoms $B$ relies on the total Hamiltonian
\begin{eqnarray}
\hat H = \hat H_A + \sum_{j=1}^N \hat H_{B_j} + \hat W + \sum_{j=1}^N \hat V_{AB_j} + \sum_{j>l} \hat V_{B_jB_l}.
\label{H}
\end{eqnarray}
Here, $H_A$ and $H_{B_j}$ are the Hamiltonians of the noninteracting atoms $A$ and $B_j$ ($j=1,\ldots, N$) which are assumed to be at rest. Taking the position of the nucleus of atom $A$ as the origin, we denote the position of the nucleus of atom $B_j$ by ${\bf R}_j$. The coordinates of the electron of atom $A$ and the electron of atom $B_j$ are ${\bf r}$ and ${\bf r}_j = {\bf R}_j + {\bm \xi}_j$, respectively ($j=1,\ldots, N$). 

In Eq.\,(\ref{H}), $\hat{W}$ is the interaction of the electrons with the external electromagnetic field. The latter will be taken as a classical, linearly polarized field, described by the vector potential 
${\bf A}= c {\bf F}_0/\omega \cos\left(\omega t - {\bf k} \cdot {\bf r}\right)$,
where $\omega=c |{\bf k}|$ and ${\bf k}$ are the angular frequency and wave vector, 
and ${\bf F}_0$ is the peak field strength. The interaction reads \cite{au}
\begin{eqnarray} 
\hat{W} &=& \hat{W}_0^+ e^{-i \omega t} + \hat{W}_0^- e^{ i \omega t} 
\nonumber \\ 
\hat{W}_0^{\pm} &=& \frac{{\bf F}_0}{ 2 \omega} \cdot \left( e^{ \pm i {\bf k} \cdot {\bf r} } \, \hat{\bf p} + \sum_{j=1}^N e^{ \pm i {\bf k} \cdot {\bf r}_j }\, \hat{\bf p}_j \right),
\label{interaction} 
\end{eqnarray} 
where $\hat{\bf p}$ denotes the momentum operator of the (active) electron in atom $A$ and $\hat{\bf p}_j$ denotes the momentum operator of the (active) electron in atom $B_j$. Below we will assume that $kR_j\ll 1$ holds for $j=1,\ldots, N$, so that we may apply the dipole approximation to the interaction (\ref{interaction}), ignoring its spatial variation and retardation effects.

The photon field may lead to direct ionization of atom $A$ or resonant excitation of one of the atoms $B$. In the latter case, atom $A$ can be ionized by radiationless energy transfer upon deexcitation via the (instantaneous) dipole-dipole coupling
\begin{eqnarray} 
V_{AB_j} = \frac{ {\bf r}\cdot {\bm \xi}_j }{ R_j^3 }
 - \frac{ 3({\bf r}\cdot{\bf R}_j)({\bm \xi}_j\cdot{\bf R}_j)}{ R_j^5 }
\label{VAB}
\end{eqnarray}
between the atoms. An analogous term $V_{B_jB_l}$ in Eq.\,(\ref{H}) describes the dipole-dipole interaction between the atoms $B_j$ and $B_l$.

In the following, all the interactions will be considered as weak and treated as small perturbations. The photoionization probability will be evaluated within second-order time-dependent perturbation theory.

The total wave function of the multi-electron system may be expanded into product states of unperturbed eigenstates of the atoms $A$, $B_1$, \ldots, $B_N$. Let, in particular, $\varphi_0(\bf r)$ and $\varphi_{\bf p}(\bf r)$ denote the ground state and continuum state (with asymptotic momentum ${\bf p}$) of atom $A$, with corresponding energies $\varepsilon_0$ and $\varepsilon_p$. Similarly, $\chi_0^{(j)}(\bm{\xi}_j)$ and $\chi_1^{(j)}(\bm{\xi}_j)$ are the ground state and first excited state in atom $B_j$, with energies $\epsilon_0$ and $\epsilon_1$ (see Fig.\,1). Applying perturbation theory up to the second order, it is straightforward to obtain the following expression for the photoionization amplitude of atom $A$:
\begin{eqnarray}
S_{\bf p}(t) = \frac{e^{i(\varepsilon_p-\varepsilon_0-\omega)t}}{\varepsilon_p-\varepsilon_0-\omega-i0}
\Bigg( \langle\varphi_{\bf p}|W_0^-|\varphi_0\rangle + \nonumber \\
\frac{\langle\chi_1^{(1)}|W_0^-|\chi_0^{(1)}\rangle}{\epsilon_1-\epsilon_0-\omega-i\Gamma_B}
\sum_{j=1}^N \langle\varphi_{\bf p}\chi_0^{(j)}|V_{AB_j}|\varphi_0\chi_1^{(j)}\rangle\Bigg)
\label{S}
\end{eqnarray}
It consists of the direct photoionization channel (arising in the first order) and the interatomic pathway (arising in the second order). $\Gamma_B$ denotes the radiative width of the state  $\chi_1^{(j)}$, and $t$ is the interaction time with the external field. The photoionization probability is obtained by taking the modulus squared of the amplitude (\ref{S}) in the limit $t\to\infty$ and integrating over the final electron momenta $\bf p$.

It is worth mentioning that photoionization of atom $A$ may, in principle, also proceed via photoexcitation of atom $B_j$ which transfers the energy subsequently to another atom $B_l$ through the dipole-dipole interaction $V_{B_jB_l}$, and only afterwards atom $A$ is ionized via ICD of atom $B_l$. In such a case, however, ionization of atom $A$ would require (at least) three interaction steps. This pathway has therefore been ignored in Eq.\,(\ref{S}) as a higher order process. Note that this is justified although the interference term of such a three-step ionization mechanism with the direct ionization channel in $|S_{\bf p}(\infty)|^2$ is of the same perturbation order as the square of the two-step interatomic channel. The latter will still be significantly larger because we consider a situation where the amplitude of the two-step channel is not only larger than the three-step channel but -- due to its resonant character -- also larger than the direct channel.

A particularly interesting situation arises when the contributions to the interatomic ionization channel of atom $A$ from the $N$ neighboring atoms in Eq.\,(\ref{S}) add fully constructively. This happens, for instance, when all atoms $B_j$ have almost the same distance $R_j\approx R_0$ to atom $A$ and are in the same plane perpendicular to the field polarization. Then, due to interference, the interatomic channel of photoionization is larger than the direct channel by the factor
\begin{eqnarray}
\rho_N \sim N^2 \left(\frac{c}{\omega R_0} \right)^6\,,
\label{rhoN}
\end{eqnarray}
where we omitted a prefactor of order unity.

\begin{figure}[t]  
\vspace{-0.25cm}
\begin{center}
\includegraphics[width=0.4\textwidth]{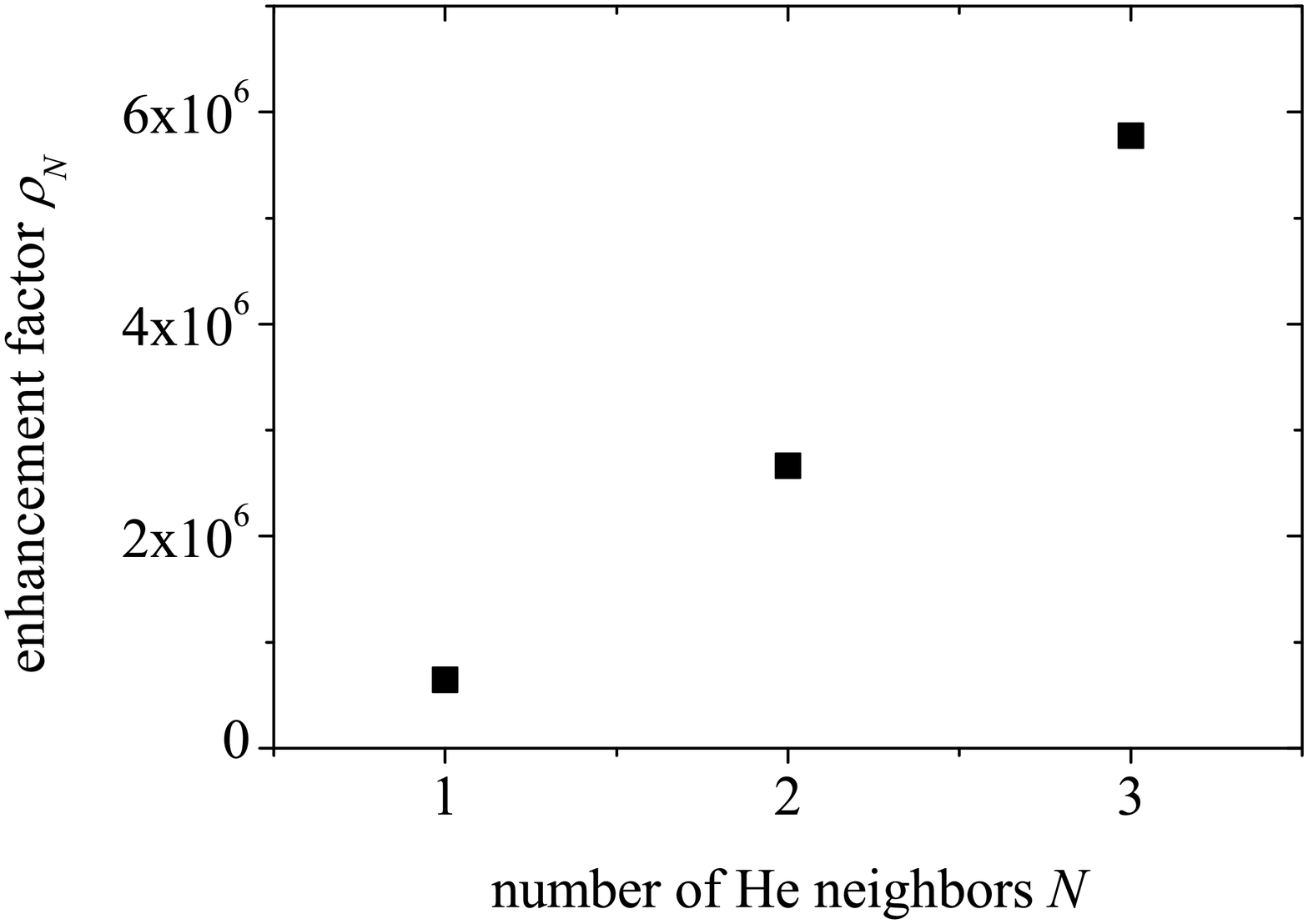}
\end{center}
\vspace{-0.5cm} 
\caption{Resonant photoionization of a Li atom in the presence of neighboring He atoms. Shown is the relative enhancement with respect to the photoionization cross section of an isolated Li atom [see Eq.\,(\ref{rhoN})], for various numbers $N$ of He neighbors. The incident photon energy is resonant with the $1s^2\,^1S$--$1s2p\,^1P$ transition in helium (i.e., $\omega \approx 21.2$\,eV). The distance between the Li atom and each of the He atoms is $R\approx $1\,nm and the He atoms are assumed to be located in the plane perpendicular to the photon field polarization. $N=1$ corresponds to the diatomic case considered in \cite{2CPI}; $N=2$ would be realized, e.g., by a linear chain He--Li--He; and $N=3$ could correspond to a Li atom in the center of an equilateral triangle formed by three He atoms.}
\label{figure2}
\end{figure}

Let us consider for illustration a simple example. The photoionization cross section of an isolated Li atom amounts to $4.5\times 10^{-19}$\,cm$^2$ \cite{LANL} at $\omega = 21.2$\,eV, which coincides with the $1s^2\,^1S$--$1s2p\,^1P$ transition in helium. When, instead, the Li atom is surrounded by, say, three He atoms at a distance of $R= $1\,nm, the photoionization is enhanced by almost seven orders of magnitude ($\rho_3\sim 6\times 10^6$) due to the resonant collective coupling between the Li and He atoms (see Fig.\,\ref{figure2}). The enhancement is so tremendous, that a very substantial effect will still remain when the finite bandwidth $\Delta\omega$ of the external field is taken into account because of which only a fraction $\Gamma_B/\Delta\omega$ of the incident photons will be resonant. Using $\Gamma_B = 4 \times 10^{-7}$\,eV and assuming $\Delta\omega\sim 100$\,meV, the relative enhancement still amounts to one order of magnitude.

The amplification by $N^2$ in Eq.\,(\ref{rhoN}) is reminescent of superradiance and related collective effects \cite{Dck,exp,bran,sono,plas,bec,moln,bald,bgm}. The present phenomenon could therefore be called ``super-enhanced photoionization.''

In a more general situation, where the interatomic position vectors ${\bf R}_j$ significantly differ from each other with respect to their length and orientation to the external field, the photoionization yield will not follow the simple $N^2$ scaling of Eq.\,(\ref{rhoN}). The photoionization cross section will rather be a complicated function of the precise ensemble geometry. Nevertheless, when all the distances $R_j$ lie in the same range, a pronounced collective enhancement of the photoelectron yield will still result, showing an $N$ scaling between linear and quadratic. Indeed, such a dependence on the cluster size $n$ has also been found in calculations of ICD of a $2s$ hole in the central atom in neon clusters Ne$_n$ \cite{NeClusters}. 

The strong enhancement of photoionization in the present study relies on the fact that the spatial extension of the multi-atom ensemble is assumed to be smaller than the (reduced) wavelength of the incident laser field. Therefore, the space-dependent factors in Eq.\,(\ref{interaction}) are immaterial and only a relatively weak spatial dependence on the ensemble geometry described by the position vectors ${\bf R}_j$ ($j=1,\ldots,N$) remains in Eq.\,(\ref{S}). Interestingly, the condition $kR_j\ll 1$ not only favors the appearance of collective effects, but also enables the dominance of the interatomic ionization channel as compared with the direct ionization channel in a system of two atoms $A$ and $B_j$ [see Eq.\,(\ref{rho1})]. 
Note that the situation in the MARPE measurements \cite{MARPE} on MnO crystals was largely different because the relevant 2$p$--3$d$ transition energy in Mn is very high, amounting to several hundred eV. In combination with the interatomic distance of about 2\AA, this leads to $kR\sim 1$.

An experimental observation of the predicted effect of collectively enhanced resonant photoionization may be possible by using an alkali atom (e.g., Li) attached to a helium cluster or water droplet. The Li photo-ionization yield could be monitored as a function of the applied field frequency in order to reveal the enhancement at the resonance. Note that experimental photo-ionization studies on alkali atoms embedded into helium and water droplets have been successfully conducted in recent years \cite{droplet, jets}. Another promising system for observation of collectively enhanced resonant photoionization might be endohedral fullerenes \cite{C60} where also the dependence of the photo-electron signal on the cluster size could be determined. 

Useful discussions with Bennaceur Najjari are gratefully acknowledged. The work was supported in part by the Extreme Matter Institute (EMMI).


\end{document}